# Revisiting, resolving and unifying the nanochannel-microchannel electrical resistance paradigm

Yoav Green[1,*], Ramadan Abu-Rjal[2,**], and Ran Eshel[2]

[1] Department of Mechanical Engineering, Ben-Gurion University of the Negev, Beer-Sheva 8410501, Israel

[2] Faculty of Mechanical Engineering, Technion−Israel Institute of Technology, Technion City, 3200003, Israel

* Email for correspondence: yoavgreen@bgu.ac.il
** Email for correspondence: ramadan.rjal@campus.technion.ac.il

**Abstract**. Until recently, the accepted paradigm was that the Ohmic electrical response of nanochannel-microchannel systems is determined solely by the nanochannel while the effects of the adjacent microchannels are negligible. Two, almost identical, models were suggested to rationalize experimental observations that appeared to confirm the paradigm. However, recent works have challenged this paradigm and shown that the microchannels contribute in a non-negligible manner, and thus these two models are inadequate in describing realistic nanochannel-microchannel systems. Two newer nanochannel-microchannel models were suggested to replace the nanochannel-dominant models. These models were limited to either very low or very high concentrations. Here, we review these four leading models. The most popular is shown to be incorrect, while the remaining models are unified under a newly derived solution which shows remarkable correspondence to simulations and experiments. The unifying model can be used to improve the design of any nanofluidic based systems as the physics are more transparent, and the need for complicated time-consuming preliminary simulations and experiments has been eliminated.

**Introduction.** It has long been thought that the Ohmic resistance of nanochannel systems [**Figure 1**(a)-(b)] behaves in a peculiar, yet rather simple-to-understand manner[1,2]. At high concentrations, the resistance, $R$, decreases linearly with increasing bulk concentration, $c_0$ (i.e. $R \sim c_0^{-1}$), while at low concentration the resistance saturates to a constant value ($R \sim const$) determined by the surface charge of the nanochannel itself [dashed red line in **Figure 1**(d)]. This behavior has been reported in numerous works[1–34] and reviews [17,35,36]. However, there is a growing body of evidence[37–41] that challenges this well accepted paradigm. Namely, it has been observed that at low concentrations, the resistance doesn't saturate but rather has a similar $R \sim c_0^{-1}$ scaling[37,39] [solid blue line in **Figure 1**(d)]. Since the Ohmic response lays at the heart of any nanofluidic-based application (electrodialysis[42–44], energy harvesting [3,34,38,45–47], fluid based electrical diodes[4,38,48–52], DNA biosensors[53–60], biomaterial modeling[33], and modeling basic physiological phenomena[61]) that span all nanoporous material (graphene oxide based membranes[5,25–27,62], carbon nanotubes[20,63,64], silicon nanochannels[6,39,65], conducting hydrogels[33], colloid based membranes[7,8], mesoporous silica films[28], exfoliated layers of a clay mineral[29], Nafion [66], cellulose nanofibers membrane from wood [30], $Ti_3C_2$ MXene membranes[9], MXene/Kevlar membranes[34], Single-layer $MoS_2$ nanopores[18], AMX-Sb anion exchange membranes[67–69] and more), it is crucial to revisit this paradigm and resolve the apparent contradiction.

Two decades ago, with the first observation[1,2] of the saturation-decrease behavior [**Figure 1**(d)], two remarkably simple models were suggested[15,16,21,24,36]. Both asserted that the total Ohmic resistance of the nanochannel-microchannel system was solely determined by that of the nanochannel and that the microchannel effects were negligible. As these models were so intuitive, they were neither challenged or scrutinized. Further, because of their inherent simplicity, they are still being used.

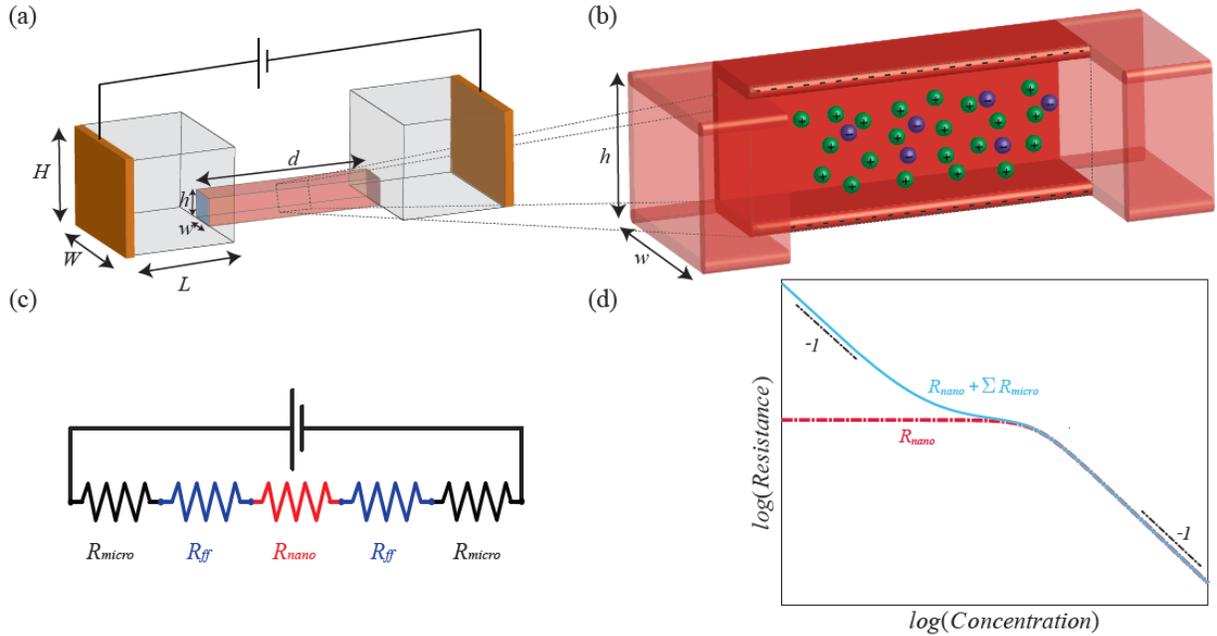

**Figure 1**. (a) A 3D schematic representation of a nanochannel-microchannel system with two electrodes at the ends connected to a power source. The nanochannel's height, $h$, is not to scale and has been exaggerated for presentation purposes. The interfaces of the microchannels and nanochannel are marked by blue and correspond to the field-focusing resistors, $R_{ff}$. (b) Zoom-up of the permselective nanochannel cross-section. Positive and negative ions are marked by green and purple spheres, respectively. Since, the surface charge is negative, there is a surplus of positive ions relative to the negative ions. (c) An equivalent electrical circuit of the system comprised of resistors in series. (d) Log-log plot of the total electrical resistance, $R$, versus the bulk concentrations, $c_0$, for a nanochannel-dominated system, $R_{nano}$, and a system accounting for both the nanochannel and microchannel resistances, $R_{nano} + \Sigma R_{micro}$. The dashed black lines denote a slope of -1.

One is then left with a set of questions: 1) Given that only one nanochannel-dominant model can be correct, which is the correct one and why is the other one incorrect? 2) Why do these models fail at lower concentrations when the effects of the microchannels are non-negligible? 3) Can the different models (nanochannel-dominant models and nanochannel-microchannel models) be reconciled? And if so, how? To answer these questions, we derived a new unifying solution (see supplement). This solution disqualifies one of the nanochannel-dominant models. It is then able to unify the three remaining models and show that they are all limiting cases of a more general solution.

To highlight the advantages of the unifying solution, we review the four leading theories and discuss their inherent assumptions, advantages, and shortcomings. Thereafter, we introduce the novel model which unifies and resolves the nanochannel-microchannel electrical resistance paradigm. We believe that the findings of this work will substantially improve the design of nanofluidic based applications, many of which are pressing to society.

To better understand the transition from an inversely linear dependence on the concentration, $R \sim c_0^{-1}$, to a saturation (i.e. concentration independence, $R \sim const$), one needs to discuss how a property termed permselectivity changes with the concentration. Permselectivity is the ability of nanochannels to filter out ions of a particular charge (positive or negative)[17,48,70]. It is controlled by the degree of the Debye length overlap of two surfaces as well as the excess counterion concentration in the channel, $N$ (which is related to the surface charge, an exact expression is given below)[52,71–73]. The degree to which a system is permselective is characterized by the transport number (described below) and ranges from two extreme cases: ideal permselectivity (low concentrations – $\tilde{N} = N/c_0 \gg 1$) and vanishing permselectivity (high concentrations – $\tilde{N} \ll 1$). How much a system needs to be permselective (or not at all) depends on the desired applications. Regardless of application, both ideal and vanishing permselective systems share a common feature: at small currents and all concentrations, the current-voltage ($I-V$) response is linear. For ideal permselectivity and larger currents, the response is no longer linear[37–39,42–44,71–76]. Herein, we focus only on the linear Ohmic resistance, $R = V/I$.

**Resistance paradigm.** For our analysis, three comments are in order. First, in the literature it is common to discuss the behavior of the electrical conductance of the system, denoted by $\sigma$. In this work we will almost solely discuss the electrical resistance, which is reciprocal to the conductance ($R = \sigma^{-1}$). The resistance formulation, as we will show, provides a more intuitive and simpler physical description. Second, throughout this work, for brevity, we will refer to the permselective region as the nanochannel. However, the 'nanochannel' could be replaced with any nanoporous material detailed above[5–9,18,20,25–30,39,62–69] and the results would still hold. Third, for the sake of simplicity, we consider here symmetric and binary electrolytes ($z_\pm = \pm 1$) with equal ionic diffusivities ($D_\pm = D$), e.g. KCl, yet our approach can be extended to any electrolyte. Hence, we emphasize that the findings of this work aren't limited to this chosen electrolyte or for a particular permselective material, but rather the results of this work are robust.

**One-layer system – superposition model**. Perhaps the most popularly used model is the one referred to here as the "superposition model" (dotted line in **Figure 2**). This model was first suggested by Schoch et al [15–17] and since then has been adopted by many[3–14,31,32,35]. It has been further extended to account for additional phenomena[18,19,41]. This model is based on two key assumptions: the total conductance is a superposition of two different states and the overall response of the system is solely determined by the nanochannel.

It is known that the electrical current passing through the nanochannel depends on both the geometry and the local conductivity, $\kappa$. This conductance, where only the bulk concentration varies, is commonly termed bulk conductance, $\sigma_{bulk}$, and is defined by $\sigma_{bulk} = 2\kappa hw/d$, where $d$, $h$, and $w$ are the length, height and width of the nanochannel, respectively [**Figure 1**(a)-(b)]. It was suggested that the total conductance is further increased by the contribution of the current transported due to the surface charge, $\sigma_s$, or excess counterion concentration $N$ which can be calculated from the requirement of overall electroneutrality $N = -2\sigma_s / Fh$ where $F$ is the Faraday constant. This contribution is termed surface conductance and is given by $\sigma_{surface} = \kappa Nhw/(c_0 d)$. Then, it is assumed that the total conductance of the nanochannel is simply a superposition of these two states

$$R_{1,super}^{-1} = \sigma_{1,super} = \sigma_{bulk} + \sigma_{surface} = \frac{hw}{\rho_{res} d}(2 + \tilde{N}), \tag{1}$$

where $\rho_{res}$ is the local resistivity and is given by $\rho_{res} = \kappa^{-1} = \Re T / F^2 D c_0$. Here, $\Re$ is the universal gas constant, $T$ is the absolute temperature, $D$ is the diffusion coefficient and $c_0$ is the bulk concentration. The subscript for the resistance, here and below, is comprised of two components, number and name. The number denotes the number of layers within the system (e.g. in this case it is one layer, only the nanochannel) and the name denotes the model (here, for brevity, we use the shortened subscript *super* to denote superposition).

*Advantages*. The model is conceptually easy to understand. Eq. (1) is simple.

*Shortcomings*. Careful inspection and comparison of $R_{1,super}$ (dotted line in **Figure 2**) with 1D simulations (markers in **Figure 2**) shows that the correspondence at the intermediate value of $N \sim c_0$, where the superposition assumption is the intended resolution, is not good. We note that there is no known rigorous derivation for Eq. (1). Rather it rests on empirical reasoning, namely, that the superposition of the bulk and surface conduction states is allowable. The failure of the model can be attributed to this assumption. Our results suggest that the superposition approach requires reexamination.

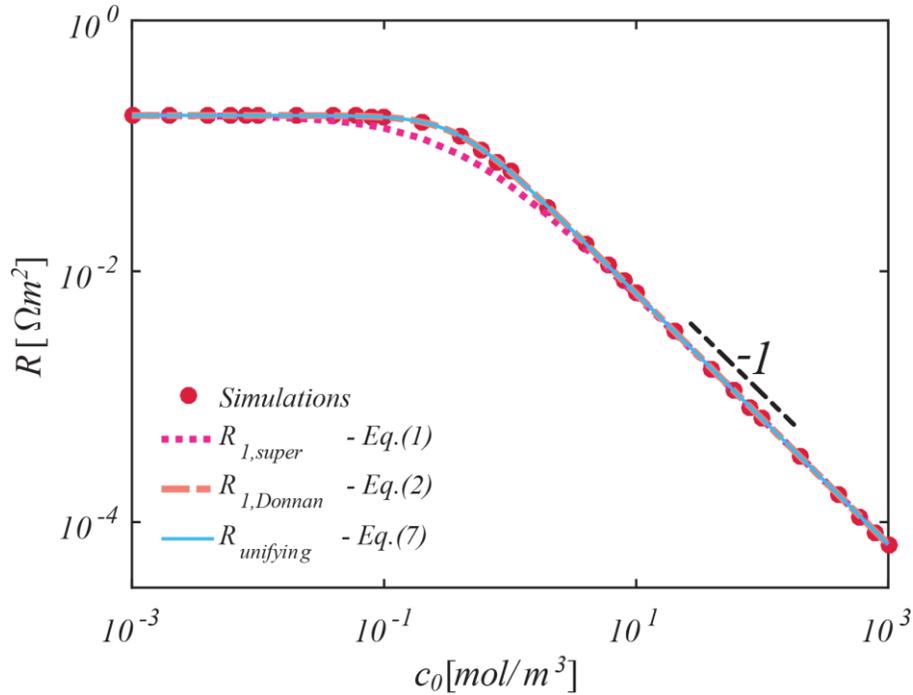

**Figure 2**. Log-log plot of the resistance, $R$, versus concentration, $c_0$, of a 1D one layered system ($L=0$ i.e. no microchannels). Dashed black lines denote a slope of -1. Simulation data taken from Ref. (39). See Supplement for simulation parameter values.

**One-layer system – Donnan equilibrium**. Like the previous model, this model also assumes that the overall response is determined solely by the nanochannel. However, here it is assumed that the nanochannel is in Donnan equilibrium with its environment. Thus, the resistance is given by[20,21,36]

$$R_{1,Donnan} = \frac{d\rho_{res}}{hw}\sqrt{4+\tilde{N}^2} \ . \tag{2}$$

*Advantages*. Eq. (2) is still conceptually easy to understand. It shows perfect correspondence to the 1D simulations for the entire range of $c_0$ (dashed line in **Figure 2**). This is unsurprising, as this model can be analytically derived from the Poisson-Nernst-Planck equations.

*Shortcomings*. Unless the electrodes are placed exactly at the nanochannel interfaces, such that microchannel length is zero, $L = 0$, the adjoining microchannels contribute additional resistors to the system[39,41]. Hence, when $L \neq 0$, this model is incapable of describing the observed additional increase of the resistance with decreasing concentration (**Figure 3** and **Figure 4**) as predicted by theory[37,38], confirmed in simulations[37,39,40] and observed in experiments[39,41,77].

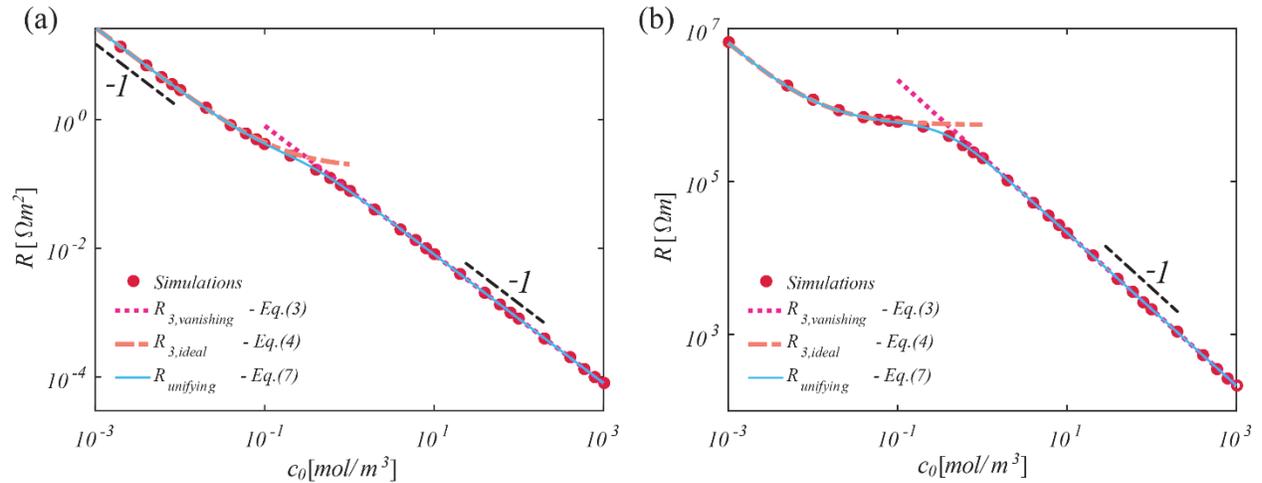

**Figure 3**. Log-log plot of the resistance, $R$, versus concentration, $c_0$, of a three-layered system comprising two microchannels connected by a nanochannel in (a) 1D (i.e. $R_{ff} = 0$) and (b) 2D. Dashed black lines denote a slope of -1. Simulation data taken from Ref. (37,39). See Supplement for simulation parameter values.

**Three-layer system – Ideal and vanishing permselectivity**. Like the previous model, this model also satisfies the Donnan equilibrium of the nanochannel. In contrast, the contributions of the adjoining microchannels to the resistance are explicitly accounted for[37–39].

Exact solutions can be derived for the two extreme cases corresponding to ideal permselectivity ($\tilde{N} \gg 1$) and vanishing permselectivity ($\tilde{N} \ll 1$). It has become common to continue to use the terminology surface and bulk conductance. However, this terminology is oversimplifying and doesn't address changes in permselectivity with changes in $c_0$. Hence, we adopt the notation of ideal and vanishing permselectivity. The expressions for the resistances are [37–39]

$$R_{3,vanishing} = \tfrac{1}{2}(R_{nano} + \Sigma R_{micro}), \tag{3}$$

$$R_{3,ideal} = \bar{R}_{nano} + \Sigma R_{micro}, \tag{4}$$

where

$$R_{nano} = \frac{\rho_{res} d}{hw}, \quad \bar{R}_{nano} = \frac{R_{nano}}{\tilde{N}}, \tag{5}$$

$$\Sigma R_{micro} = 2(R_{micro} + R_{ff}), \quad R_{micro} = \frac{\rho_{res} L}{HW}, \quad R_{ff} = \frac{2\rho_{res} f}{L}. \tag{6}$$

Here $L$, $H$, and $W$ are the length, height and width of the microchannels, respectively [**Figure 1**(a)]. The resistance attributed to the microchannel regions, $\Sigma R_{micro}$ [Eq. (6)], is comprised of two terms[37–39]: 1) a resistance attributed to the rectangular geometry of the system, $R_{micro}$; 2) a resistance attributed to the focusing of field lines from the larger microchannels into the smaller nanochannel, $R_{ff}$ (see supplement for an exact expression for $f$). In the main text, we consider the case where the microchannels are identical. This need not be the case and a more general expression for $\Sigma R_{micro}$ is given in the Supplement.

The transition of $R_{3,vanishing}$ to $R_{3,ideal}$ with decreasing concentration is attributed to the variation of the counterion transport number, $\tau$. The transport number, defined as the ratio of the flux of the counterions to the total current density, is a major transport characteristic of a permselective system. For an ideal binary system, only counterions are transported through the nanochannel, such that their transport number equals unity ($\tau_{\tilde{N}\gg1}=1$) and the nanochannel resistance, $\bar{R}_{nano}$, depends on $N$ yet is independent of $c_0$. For a non-ideal system, the counterion transport number lies between one-half and unity, $\tau \in [0.5,1]$. At the other extreme of vanishing permselectivity ($\tilde{N}\ll 1$), the transport number equals one-half ($\tau_{\tilde{N}\ll 1}=\tfrac{1}{2}$) and the nanochannel resistance, $R_{nano}$, depends inversely on $c_0$ yet is independent of $N$. The change of the nanochannels permselectivity (i.e. the transport number) also changes the transport characteristic in the microchannels. When both charge carriers are transported through the nanochannel the resistance decreases by a factor of 2 as is evident in $R_{3,vanishing}$.

*Shortcomings*. Each model is limited to either $\tilde{N}\ll 1$ or $\tilde{N}\gg 1$. As can be expected, the solution in the $\tilde{N}\sim 1$ region, where the two models intercept (**Figure 3**), is only approximate.

*Advantages*. The advantages are numerous. First, the derivation of the exact solution from the Poisson-Nernst-Planck equations for the two extreme cases, ideal and vanishing permselectivity, is relatively simple and straight-forward[37–39]. Second, while in this work we only consider the Ohmic response, the solution for ideal permselectivity also holds for the limiting current, $I_{lim}$, regime where the $I-V$ response is no longer linear[38,52]. Third, and perhaps the greatest advantage, these models have excellent correspondence with simulations [1D –**Figure 3**(a) and 2D – **Figure 3**(b)] and 3D experiments[39] [**Figure 4**]. Finally, these models provide a very clear and simple physical picture - the total resistance of the system is that of a series of resistances [Eqs. (3)-(4), **Figure 1**(c)]. In contrast to the resistance formulation where the removal/addition of one component/resistor is intuitive or straightforward, in the conduction formulation $(\sigma = R^{-1})$, this is no longer the case. This is why the resistance formulation is preferable.

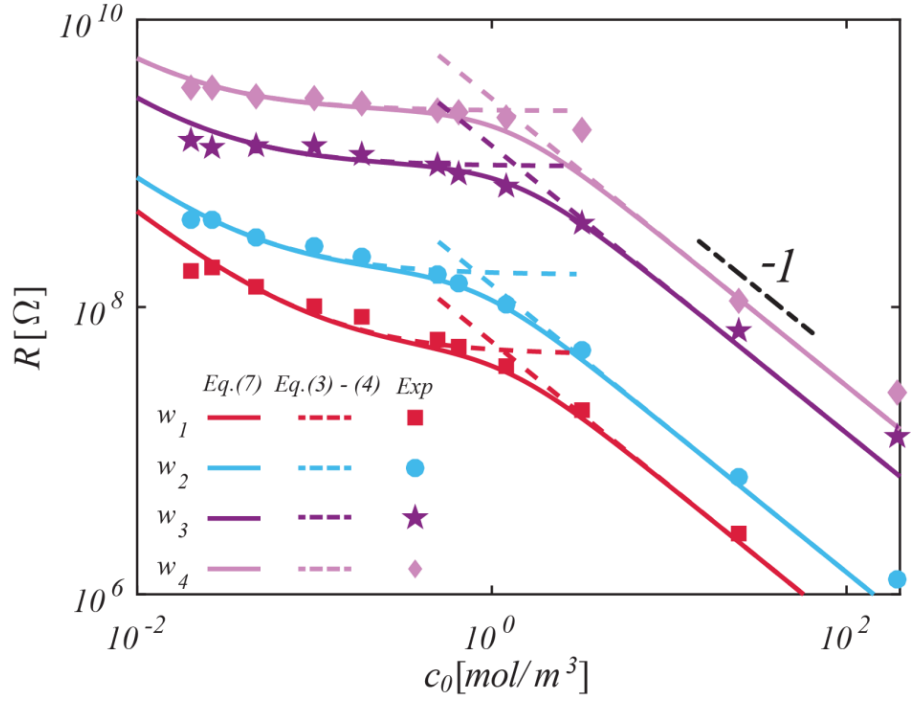

**Figure 4**. Log-log plot of the resistance, $R$, versus concentration, $c_0$, of 3D experiments of a three layered system for four different nanochannels of decreasing width $(w_1 > w_2 > w_3 > w_4)$. Dashed black lines denote a slope of -1. Experimental data is taken from Ref. (39) and the experimental conditions are given in the Supplement.

**Three-layer system – unifying solution**. Here we present a newly derived analytical expression for the Ohmic response of a 3D three-layer system which holds for all values of $N$ and $c_0$ (see Supplement). Our derivation is based on the recent method proposed in Ref. (74) that solved for the $I-V$ using the Donnan equilibrium formulation. Ref. (74) considered the general $I-V$ response and, as a result, their final equations are transcendental equations and require numerical evaluation. However, it appears that they overlooked an existing analytical solution within their complicated transcendental equations corresponding to the Ohmic response. Here, we present that solution

$$R_{unifying} = (2\tau - 1)\bar{R}_{nano} + \frac{1}{2}\left[1 + (2\tau - 1)\sqrt{4\tilde{N}^{-2} + 1}\right]\Sigma R_{micro}, \tag{7}$$

and the transport number is given by

$$\tau = \tfrac{1}{2} + \tilde{N}\left(2\sqrt{4 + \tilde{N}^2} + 4\frac{\Sigma R_{micro}}{R_{nano}}\right)^{-1}. \tag{8}$$

**Advantages**. To the best of our knowledge, this is the first analytical expression for the two most essential properties of a permselective system, the Ohmic resistance [Eq. (7)] and the transport number [Eq. (8)], that depends on the entire geometry (nanochannel and microchannels) for all values of $N$ and $c_0$. Remarkably, the expressions for the transport number and the Ohmic resistance are a simple function of two ratios: $\tilde{N} = N/c_0$ and $\Sigma R_{micro}/R_{nano}$. Thus, given the geometry and the surface charge, the transport and permselectivity can be easily calculated given the concentration. Further, Eq. (7) reduces to Eqs. (2)-(4) at each appropriate limit (see Supplement) and thus unifies these three models. The correspondence of Eq. (7) for all values of $c_0$ and geometries in 1D-2D simulations (**Figure 2**-**Figure 3**) and 3D experiments (**Figure 4**) is outstanding.

**Shortcomings**. Albeit a conceptual complication, described below, there are no shortcomings.

Eqs. (3) and (4) suggest that in the limiting cases of vanishing and ideal permselectivity, the response could be described by a simple electrical circuit. This is not apparent upon substitution of Eq. (8) into Eq. (7) whereby the resultant expression is complicated and depends non-linearly on $\Sigma R_{micro}$ and $R_{nano}$.

To circumvent this issue, we suggest not to view the final result as a single equation, which perhaps over-simplifies the assumption. We note that, in Eqs. (3) and (4), the transport number was enforced a priori, and is thus a hidden equation, from which, as a consequence, the resistance is determined. Thus, if Eqs. (3) or (4) were to be presented in a detailed manner, then each of their according transport number needs to be presented. Hence, the resolution is to view $R_{unifying}$ and $\tau$ as two separate variables, and, in this manner, the analysis is indeed as simple as the previous models. Accordingly, Eq. (7) can be treated as a series of weighted resistors that are connected together, whereby the weight is determined by the transport number [Eq. (8)]. Thus, the only apparent shortcoming has been circumvented and we are left to conclude that this model doesn't have any apparent shortcomings.

**Conclusions.** This work focuses on the Ohmic resistance of a permselective nanochannel flanked by two microchannels. We reviewed the shortcoming and advantages of four leading models whose properties are outlined in **Table 1**. We showed that one of these models is incorrect, and the remaining three models, while correct, suffer from certain limitations. To bypass these limitations, we reduce the number of embedded assumptions in the derivation and derived a novel model that unifies these three models. The resultant expressions, given by two equations [Eqs. (7) and (8)], appear to correspond perfectly to both simulations and experiments for all geometries and for all concentrations.

This novel model [Eqs. (7) and (8)] holds for any nanoporous material used for electrokinetic transport of ions and provides analytical expressions for the two most essential properties of a permselective system. It can be used to assess what is the transport number, as well as the relative contribution of the permselective material to that of the adjoining microchannels, which are prevalent in realistic systems, without reverting to time and energy consuming numerical simulations or experiments that can be highly complicated and require much skill. Most importantly, the physics of ion transport through nanoporous materials is clearer and easier to understand.

Table 1. Summary of assumptions, advantages and limitations of all models.

|  | Derivable[1] | Geometry[2] | Concentration[3] | Current[4] | Sim/Exp[5] |
|---|---|---|---|---|---|
| $R_{1,super}$ | No | $L=0$ | All $c_0$ | $I \ll I_{lim}$ | No |
| $R_{1,Donnan}$ | Yes | $L=0$ | All $c_0$ | $I \ll I_{lim}$ | Yes |
| $R_{3,ideal}$ | Yes | None | $N \gg c_0$ | $I_{lim}$ | Yes |
| $R_{3,vanishing}$ | Yes | None | $N \ll c_0$ | $I \ll I_{lim}$ | Yes |
| $R_{unifying}$ | Yes | None | All $c_0$ | $I \ll I_{lim}$ | Yes |

[1] Model can be derived from Poisson-Nernst-Planck equations.
[2] Geometric limitations.
[3] Concentration range limitations.
[4] Current range limitations (Ohmic response - $I \ll I_{lim}$, Limiting current - $I_{lim}$).
[5] Good correspondences with simulations and/or experiments.

**Model Derivation.** Details are given in the Supplementary.
**Numerical Simulations.** Data taken from Refs.(37,39). See Supplementary for simulation values.
**Experimental Data**. Data taken from Ref.(39). See Supplementary for geometric values.
**Author Contributions.** Y.G. conceived the idea, derived the solution and conducted the analysis. All authors discussed the material and wrote the manuscript.
**Notes.** The authors declare no competing financial interest.